\documentstyle[12pt,aps,epsf]{revtex}

\input{epsf}
\begin{document}

\title{Chasing 'Slow Light'}
\author{E.B.Aleksandrov$^*$ and V.S.Zapasskii$^{**}$}
\maketitle
 \hskip20pt

\vskip10pt
\centerline{ $^*$Ioffe Physico-Technical Institute, St. Petersburg, 194021 Russiaa}
\centerline{ $^{**}$All-Russia Research Center ''Vavilov State Optical
Institute'',}
\centerline{St. Petersburg, 199034 Russia;}

\centerline{St. Petersburg State University, Institute of Physics, St. Petersburg, 1985044
Russia}
\centerline{ {\it  e}-mail:  zap@vz4943.spb.eduu}
\vskip20pt
\begin{abstract}
A critical review of experimental studies of the so-called 'slow light' Arising from
 the anomalously high steepness of the refractive index dispersion under
 conditions of electromagnetically induced transparency or coherent population
 oscillations is presented. It is shown that a considerable amount of experimental evidence for observation of the 'slow light' is not related to the low group velocity of light and can be easily interpreted in terms of a standard model of interaction of light with a saturable absorber.
\end{abstract}

\vskip20pt

\normalsize

\section{INTRODUCTION}

In the last decade, the topic of 'slow' and 'fast' light - light pulse
propagating in a medium with ultralow, superluminal, or even negative group
velocity - has gained particular popularity. The keen interest in this topic is
not primarily related to the possibility itself of controllably varying the
light group velocity in a medium, but instead to the a huge scale of these
variations (7 orders of magnitude and more) and to the claimed favorable
prospects for the application of 'slow light' in telecommunication and optical
computing. As a result, in the late 90s, the 'slow light' has, in fact, turned
into a separate trend of physical optics. Several hundreds of publications have
been devoted to these problems. They are discussed on special topical meetings
and workshops. A somewhat sensational character of the claimed achievements
(including the effects of 'stopped' and 'stored' light) renders this topic
popular not only in scientific literature, but also in mass media. Concurrently
with the increased rate of publications on 'slow light' research, the quality of
the publications was getting noticeably poorer. The phenomenological simplicity
of those manifestations of 'slow light' made it possible to observe similar
effects in relatively simple experimental conditions and ascribe them to
'slowing down' and 'stopping' of light without sufficient grounds. In these
notes, we critically analyze the publications on 'slow light' and demonstrate
inconsistencies in a considerable number of the arguments intended to provide
evidence for dramatic changes of the light group velocity.

\section{FROM THE HISTORY OF GROUP VELOCITY OF LIGHT}

The problem of slowing down of light in a medium has a long history, dating back
to Newton's 'Optics'. Early in the last century, attention was again attracted
to the problem of the velocity of light in a medium in connection with the
as-developed special theory of relativity, which limited (in its second
postulate) the speed of transfer of information by the velocity of light in
vacuum (c). The propagation of electromagnetic waves in real media was studied
in the early 20th century by founders of physical optics. It is worth mentioning
A.Sommerfeld and his disciple L. Brillouin, who summarized the results of these
studies in a monograph [1]. Later on, the velocity of light in a medium
attracted the attention of researchers, mainly, in the cases when it strongly
differed from that in vacuum. Forty years ago, a 'superluminal' group velocity
of light was claimed to be observed in [2]. The studies were performed with the
light pulses propagating in a gain medium under conditions of strong
nonlinearity related to the gain saturation under the action of the pulse. In
this case, the shape of the pulse changed as it propagated through the medium,
with its peak being shifted in the forward direction. By defining the group
velocity of light as the velocity of motion of the pulse maximum, the authors of
[2] claimed observation of the 9-fold excess of the velocity of light in the
medium over that in vacuum. This phenomenon was, in essence, the inverse effect
of the light pulse delay in a saturable absorber (see, e.g., [3]) and,
therefore, was not of any principal interest . \footnote{ It is noteworthy that
neither in this paper, nor in any other subsequent report on superluminal
propagation of the light pulse, do the authors claimed to refute the special
theory of relativity (see, e.g., review [4]).}

A new splash of attention to this issue was associated with the {\it linear}
propagation of optical pulses whose spectrum fell into the region of steep
dispersion of the refractive index of a medium. In this case, in the standard
formula for the light group velocity $V_g$

\begin{equation}
V_g={c\over n+\omega \hskip0.5mm dn/d\omega}
\end{equation}

the dispersion term starts to dominate. As a result, the motion of the light pulse in
the medium slows down not so much due to the deviation of the refractive index n
from unity (the resources of this mechanism are rather limited), but mainly due
to steep dependence of the refractive index on the frequency $\omega$. This
dispersion-related contribution affects only the group velocity
of light and does not change its phase velocity controlled by the value of $n$.

It is noteworthy that the problem of definition of the light group velocity
frequently becomes the subject of controversy. The notion of group velocity
introduced in \cite{1} is fairly definite for a bi-frequency wave, being defined
as the speed of motion of the beat envelope. This, rather abstract, field is
evidently of limited interest in practice. As applied to the light pulse with a
continuous spectrum, the problem of group velocity becomes more complicated,
since the pulse, while traveling through a dispersive medium, changes its
shape, and the quantitative evaluation of its displacement in space becomes
ambiguous. In what follows, we will understand the light group velocity as it is
defined by Eq. (1). For transparent media, this definition corresponds to the
velocity of motion of the pulse maximum and, in a certain approximation, with
the velocity of motion of the pulse as a whole. It is known, for instance, that
if the spectrum of the pulse fits into the region of linear dispersion of the
refractive index, the original pulse of Gaussian shape will propagate through
the medium practically without any distortion (see, e.g., \cite{5}), provided
that the absorption (gain) of the medium does not strongly change within the
spectral width of the pulse (see also the experiment \cite{6}).

Thus, the group velocity in the sense of Eq. (1) is the speed of motion of the
pulse maximum, but it does not mean that the opposite statement is also valid,
i.e., that the speed of motion of the pulse maximum is always the group velocity
of light. Definition (1) is applicable to a {\it transparent, linear}, and {\it
stationary medium}, whose properties change, neither due to external perturbations, nor
due to self-action of the light pulse, in the process of the pulse
propagation through the medium(i.e., the medium is supposed to be {\it linear} with respect to
the probe pulse whose group velocity is measured). For the absorbing, nonlinear, and nonstationary
media, there may exist other mechanisms of the pulse maximum delay (e.g., of the
time vignetting type), which have nothing to do with the dispersion of the
medium and cannot be described by Eq. (1). In our opinion, the delays of this
kind cannot be attributed to changes in the light group velocity in the commonly
accepted understanding of this notion.

\section{'SLOW LIGHT' AND THE GOAL OF THESE NOTES}

A real boom of publications devoted to the anomalies in the light group velocity (or
to 'slow light'
  \footnote{ In what follows, for brevity, we will  consider 'slow light' to be
   all the phenomena associated with dramatic changes of the light group
   velocity due to steep dispersion of the refractive index, i.e., the
   'slow light', 'fast light', and the light with 'negative' group velocity.}
)
was associated with realization of the electromagnetically
induces transparency (EIT) \cite{7} which has made it possible to combine two,
previously exclusive, properties of the medium - a high steepness of dispersion
and transparency. The effect is usually observed on the systems with the
energy-level diagram schematically shown in Fig. 1 (the so-called $\Lambda$-scheme). Two
electromagnetic fields, $\omega$ñ and $\omega$, are
applied to transitions between two, usually closely spaced, sublevels of the
ground state ($|1\rangle$ and $|2\rangle$) and the third common excited level $|3\rangle$. The
excitation of this type is capable of creating coherence between the two lower
states. If the frequency $\omega$ of one (usually stronger) field is fixed, and
the frequency $\omega$ of the other (probe) field is being scanned, then, in the
point of the two-photon resonance, one will detect a dip in the absorption
spectrum ('transparency window'), corresponding to a destructive quantum
interference of the transitions to the excited state via two alternative routes.
To this dip, in accordance with the Kramers-Kronig relations, should correspond
a region of steep dispersion of the refractive index. This steepness may be
extremely high due to extreme narrowness of the coherence resonance of the ground
state sublevels. Under these conditions, it appears possible to demonstrate
anomalously high retardation of a light pulse in the medium \cite{8,9,10}. The most
impressive results with the reduction of the light group velocity by more than seven
orders of magnitud have been obtained using ultracold sodium atoms (in the
vicinity of the Bose-Einstain condensation temperature) \cite{11,12}.

In more recent experiments, the circle of mechanisms capable of providing 'slow light'
was widened. The studies were performed in rather simple experimental
conditions, and analysis of the results was made, in our opinion, not critically
enough: Any apparent delay of the light pulse in the medium was attributed to its
low group velocity. Some obviously erroneous works of this type were criticized
in \cite{13,14,15}. However, the essence of the matter, in our opinion, consists
not in the errors committed in certain papers, but rather in the inconsistencies
of the provided evidence. As a result, there are presently doubts about the reliability of the
conclusions drawn by considerable number of publications on 'slow light'.

In these notes, we consider some effects of {\it incoherent} nonlinear optics
which are phenomenologically similar to the 'slow light' effects and should be
  observed in slow relaxing media with nonlinear absorption, while having
nothing to do with 'slow light'. We attract attention to a number of commonly
accepted pieces of evidence for 'slowing down' and 'stopping' of light that
cannot be used as such. We also discuss certain specific features of the 'slow light' effects
 that allow one to experimentally distinguish them from the effects of 'slow
 medium' which are
universally observable in resonant media with saturable absorption.

Since the problem of interpretation of experimental observations considered in
these notes are of a conceptual nature, and the effects of incoherent nonlinear
optics that mimic the 'slow light' are well known  and have been
repeatedly described in the literature, we consider it possible to restrict
ourselves to a qualitative level of their treatment without reproducing the
known theoretical derivations (which, if needed, can be found in the
references).

\section{THE SUBJECT OF STUDY -A NONLINEAR ABSORBER}

The simplest (and chronologically the first discovered) effects of nonlinear
optics were related to resonant perturbation of the populations of quantum states
(effects of saturation) \cite{16}. For sufficiently long relaxation times, these
{\it incoherent} nonlinear effects can, in principle, be observed at arbitrarily
low light intensities. The opposite side of the high nonlinearity of such 'slow'
systems is a certain sluggishness of their optical response and, therefore,
their limited applicability to the 'state of the art'  broadband systems of optical
information processing. These effects, as a whole, are naturally considered to
be well known and, perhaps, for this reason, are not given sufficient attention.

The effects of saturation, in their simplest form are known to be revealed when
the time of interaction of the light with the nonlinear medium substantially
exceeds the transverse relaxation (dephasing) time of the resonance oscillators
and when the rate of the light-induced transitions becomes comparable with that
of the population relaxation. In this case, the populations of the system
and, hence, the absorption of the medium change. Most frequently, the absorption
of the medium drops with increasing light intensity (the medium is {\it bleached}).
But there also exist situations when absorption of the medium increases with
light intensity. The appropriate nonlinear absorbers are called {\it inverse}.
For practical applications (mainly in laser techniques), saturable absorbers
are used in a diverse array of optical media, like dye solutions, doped crystals and glasses,
dielectrics with semiconductor nanocrystals, etc. A typical 'slow' saturable
absorber is the ruby crystal, where the coherence of the light-induced excitation
of the medium in the blue-green
spectral range is rapidly destroyed by the fast relaxation from the excited state of
the Cr$^3+$ ions to metastable levels, whose long lifetime provides retardation of
the optical response of the crystal.

The most perfect model of saturable absorber is, however, provided by the
optically pumped atoms of alkali metals, which are also characterized by fairly
long ground-state population relaxation times, but, in addition, show one more
important property: in the absence of an external magnetic field, their
anisotropy is entirely controlled by the anisotropy of the acting light (see, e.g.,
\cite{17}). As a result, in the field of polarized light, these systems may be considered as
{\it polarization} saturable absorbers, which differ from the conventional ones
by additional degrees of freedom - polarization of the exciting light and
anisotropy of the medium. These degrees of freedom considerably widen the range
of phenomena observed under conditions of optical pumping.

Below, we will dwell, in more detail, upon the properties of these 'slow' saturable
absorbers, which are basic subjects of the 'slow light' experiments. We
will consider the effects that should necessarily be observed in 'slow'
nonlinear absorbers regardless of spectral features of the medium's dispersion
and variation of the light group velocity. All the effects considered below are,
in fact, combinations of elementary properties of a saturable absorber and do
not contain anything essentially new. Still, it is exactly these effects that
are frequently considered as manifestations of 'slow light'.

\section{ON CAPABILITIES OF SATURABLE ABSORBER}

So, the question is what is to be expected from a saturable absorber under
conditions of optical excitation typical for the 'slow light' experiments and
what really new has been found in those experiments?

{\bf 1. Time-domain response.}
 The simplest property of the saturable absorber is related to the dynamics of
its response to a change in the incident light intensity. When the light
intensity at the entrance changes in a step-wise way, the light intensity at the
exit, evidently, experiences a jump, corresponding to the linear light
transmission, and an exponential growth (or exponential fall for the inverse
saturable absorber), corresponding to the process of establishing a
steady-state populations of the system (Fig. 2).

Experimental dependences of this kind are reported, for instance, in \cite{18},
where the probe-pulse shape distortion is ascribed to the effect of steep
dispersion of refractive index of the medium in the region of resonance of the
electromagnetically induces transparency and electromagnetically induced
absorption. These temporal dependences, as being of no interest, usually are not
presented in the papers on 'slow light'. We mention this type of response only
for completeness sake and for passing to the next, more constructive, item.

{\bf 2. Frequency-domain response}
 Fourier-transform of the above time
dependences yield their frequency-domain representation, which can be easily
obtained experimentally by measuring the frequency dependence of the amplitude
 and phase of oscillations of a modulated light beam transmitted through a
 saturable absorber. Figure 3 shows what the dependencies of this kind look
 like for a bleachable absorber. As expected, the frequency dependence of the
 response displays  a peculiarity in the range of low frequencies comparable
 with the inverse population relaxation time of the absorber ($\tau$). Absolute time
 delay $\Delta t(\omega)$ of the intensity modulation signal is seen to be the greatest in
 the region of lowest frequencies, while the relative (i.e., phase) delay
 $\varphi(\omega)$
 reaches its maximum at frequencies $\omega\sim 1/\tau$ and does not exceed a small fraction
 of the oscillation period.

In \cite{19,21,22,23,24}, these simple properties of nonlinear absorbers
(crystals with paramagnetic impurities) have been erroneously ascribed to the
reduction of the group velocity of light due to spectral hole burning under
conditions of coherent population oscillations. Full agreement of the
experimental data obtained in those papers with the prediction of the simple model
of saturable absorber was shown in \cite{15,25}, where one can find a more
detailed and grounded criticism of these publications \cite{19,20,21,22,23,24}.
The same perverse interpretation of the effects of delayed
photoresponse in a saturable absorbed has received further development in the
experiments with bacteriorhodopsin molecules in a polymer film \cite{26}.

{\bf 3. The light pulse delay} When a light pulse travels through a saturable absorber, the
absorption of the medium, in the general case, changes in the process of its
propagation. This leads not only to a change of the pulse amplitude, but also to
a distortion of the pulse shape. If the shape of the pulse is smooth and its width is
comparable with the relaxation time $\tau$ or exceeds it, then the distortion of its
shape appears to be small, and, in the first approximation, is reduced to a pure
shift in time. Note that the sign of the delay is positive for the usual
bleachable absorber and negative for the inverse saturable absorber. This effect
was studied as far back as 60s of the last century (see, e.g., \cite{3,27}). In
the same category of the effects it can also be attributed the effect (already mentioned
above) of apparent increase of the light group velocity in a 'saturable
amplifier' \cite{2,28}, when, like in the inverse saturable absorber, the pulse
is distorted in favor of its front edge. Figure 4 shows the calculated curves
that illustrate the pulse 'delay' for the usual (a) and inverse (b) saturable
absorber for some ratios of the pulse width d and relaxation time t \cite{15}.
Note that the illusion of the pulse delay arises, in this case, due to the
amplitude normalization of the output pulse. With no normalization, the output
pulse, in this figure, would always lie inside the input one. The light pulse
delay in the saturable absorber, evidently, is not related to the dispersion of
the refractive index. This delay is a consequence of the light-induced
nonstationarity of the medium and, in essence, is not a delay as such. In the
already mentioned papers on hole-burning under conditions of coherent population
oscillations, this delay of the light pulse, with no additional justification,
was ascribed to 'slow light' (see, e.g., \cite{2,26}).

{\bf 4. Dynamics of polarization response}
 Under the action of a resonant polarized
light, the polarization saturable absorber, mentioned above, becomes dichroic.
In accordance with the general laws of symmetry (Neumann's principle), the type
of the light-induced dichroism (linear, circular, or elliptical) should correlate
with polarization of the acting light. As is known, these properties are displayed,
in particular, by the optically pumped atomic systems, in zero magnetic field,
amenable both to orientation and alignment (this fact was recently demonstrated
once again in \cite{29}). In the case of a bleachable absorber, the light-induced
dichroism of the medium corresponds to its bleaching in the polarization of the
acting light. When polarization of the incident light is changed, the anisotropy
of the nonlinear medium and the light polarization at the exit of the medium follow
these changes with a certain time delay controlled by the characteristic time of the
absorber (for more detail, see \cite{14}).

A pulse of polarization modulation can be formally represented as a pulsed
admixture of the orthogonally polarized component to a polarized beam . (in
practice, this is usually made by phase shifting the polarization components in a
single beam with the aid of polarization modulators like, e.g., a Pockels cell).
Polarization dynamics of the pulse at the exit of the medium can be also
monitored by detecting this orthogonal component of the output beam (Fig. 5).

Figure 6 shows what the weak pulse of the orthogonally polarized component looks
like at the exit of a saturable (bleachable) absorber for some ratios between
the pulse width $\delta$ and relaxation time $\tau$. As is seen from the figure, the
distortion of the pulse shape at the exit (as for the case of pure intensity
modulation, see item 3) decreases with increasing pulse width, gradually
approaching pure shift. For the bleachable polarization absorber, this shift is
positive (delay), and for the inverse absorber, negative (advance). In this
configuration of the experiment, in contrast to the case of a 'single-channel'
scheme described in item 3, the observed delay of the polarization component, as
one can see by comparing Figs. 4 and 6, may be fairly large. Additionally, in this case the delays, although not apparent, is nevertheless real in the sense that the signal at
the exit can be observed after completion of the input signal (due to permanent
presence of the initial polarization component transformed by the perturbed
nonlinear medium).

The nature of the above delay of the polarization pulse in a 'slow' nonlinear
absorber evidently reflects only the retarded dynamics of the medium and has
nothing to do with variation of the light group velocity in terms of Eq. (1).
However, in the works on 'slow light', the delay of this kind is usually
ascribed to the group velocity reduction (see, e.g., \cite{30,31}). Note also
that the above scheme with admixing of the orthogonal polarization component and
its separation at the exit (Fig. 5) makes it possible to monitor the
polarization dynamics of the light at the exit of the absorber, but does not
allow one, by any means, to {\it probe} the medium by this polarization
component no matter how weak the probe. This is because the anisotropy of the absorber,
under the action of the polarization pulse, varies in time, and the initial
polarization components cease to be independent (normal) modes. Therefore, the
delay of the pulse of one polarization component, in this experimental
configuration, cannot be ascribed to the low group velocity of light for the
additional reason that the notion of velocity for the waves that are not normal
cannot be introduced \cite{32}.

{\bf 5. Intensity-related characteristics of the response}
 At low light intensities,
the relaxation rate of the saturable absorber (exponential regions in Fig, 2) is
intensity-independent and is determined by the 'dark' population relaxation time
T1. With increasing intensity, the effective population relaxation rate (t-1),
controlled by the sum of the dark and light-induced relaxation processes, grows
in a linear way (see, e.g., \cite{16}), whereas the delay time of the pulse or
of the light intensity oscillations (at sufficiently low frequencies, see item
2), correspondingly decrease hyperbolically. For this reason, the dependence
presented in Fig. 7 demonstrate standard properties of a saturable absorber and
cannot be used to confirm the 'slow light' model, as this is made in \cite{33}.

{\bf 6. 'Storage' of the light-induced anisotropy}
 The real delay of the polarization
pulse at the exit of a nonlinear absorber with respect to the input pulse
signifies, in particular, that the contribution to the anisotropy of the medium,
induced by the polarization pulse, persists for some time after the end of the
pulse. This time is evidently determined by the relaxation parameters of the
medium, which, as was already pointed out above, depend on the light intensity. For
this reason, by switching off the light beam immediately after completion of the
polarization pulse, we can 'freeze' the residual
anisotropy induced by the polarization pulse and 'store' it during the dark
relaxation time T1, which may be rather long. If, after such a pause, we again
switch the light beam on, then, to within the relaxation during the pause, we
will reproduce the situation existed at the moment of switching the light off.
It means that the light beam will continue 'readout' of the anisotropy of the
medium induced by the polarization pulse, and at the exit of the medium we will
detect the 'tail' of this process (Fig. 8). The dependence of the relaxation
rate on the light intensity will be evidently revealed in the fact that the
length of this 'tail' will vary with the pump beam intensity (see, e.g.,
\cite{34}). In our opinion, this natural manifestation of the relaxation properties
of a polarization nonlinear absorber may be considered as the effect of
polarization memory, but there are no grounds to consider it as 'stopped light'
as it is done in many papers on 'slow light' where a degenerate $\Lambda$-scheme
is used (see., e.g., \cite{30,31,35,36,37}; for more detail see \cite{13,14}).

{\bf 7. Polarization-interference paradoxes}
 Let us turn again to the polarization scheme shown in Fig. 5 where the polarized
light passing through a nonlinear absorber composed of two mutually
coherent and orthogonal polarization components is analyzed, at the exit of
the medium in the same polarization basis. Let these components, for definiteness, be
linearly polarized and in-phase, so that being superimposed, they form the light of
linear polarization. Consider the case where, like in the effect of
electromagnetically induced transparency (EIT), one of the beams (the pump or coupling
beam) is strong, whereas the other (probe) beam is weak. Then we come to the effect
which may seem, at first sight, paradoxical and which is widely exploited in the
'slow light' studies.

Under the action of only the strong pump beam, the medium is bleached and the
beam propagates over its own polarization route (C-C in Fig. 5). The weak probe
beam in the absence of the pump, also propagates along its own route (P-P),
but, being unable to bleach the medium, will experience strong absorption and
appears to be strongly attenuated at the exit. However, if we measure
transmission of the probe light {\it in the presence of the pump}, we will find
that the system is transparent. Therefore, in this
experimental arrangement, bleaching of the medium in one polarization makes it
bleached for an arbitrarily weak beam of orthogonal polarization. Paradoxicality of
this result is that the probe beam, as it may seem, should be a {\it normal}
wave of the medium whose anisotropy is formed by the beam of orthogonal
polarization and should be able to {\it probe} the medium in conformity with all the laws of
polarization optics.

Of course, there is no riddle here. On the one hand, this is a simple
combination of nonlinear absorption with an effect due to the interference of polarized
rays. If we trace the {\it amplitudes} of the two fields (it is exactly what
should be done when analyzing interference effects), we will see that addition
of the probe-field component gives rise to a slight rotation of polarization plane of the
acting field. Due to high intensity of the field, the anisotropy (dichroism) axis
of the medium follows this rotation virtually with no delay, and the medium
remains bleached for the new polarization of the pump. As a result, the beam
(and each its component) passes through the medium with no attenuation. Keep in mind that
this is a sort of play on words. We tacitly assumed that the weak
beam {\it probes} the medium, while, as was already pointed out, this is not
correct (to really probe the medium by this beam, it suffices to switch the pump
off; we will see that the medium is opaque).  Anurthermore, it will eventually be correct
to say that this is the EIT effect or
what this effect turned into in the degenerate $\Lambda$-scheme. However, whereas for
unequal frequencies of the beams in the non-degenerate $\Lambda$-scheme, with the
measuring time being much longer than the beat period, the scheme in Fig. 5 is
consistent in the sense that the probe beam, corresponding to a normal wave of
the medium, is truly able to {\it probe} it. However for equal frequencies this
approach is unphysical. For this reason, in the degenerate polarization
$\Lambda$--scheme, the notions of the 'probe' and 'pump' beams are improper. These
reasoning refers to all the papers on 'slow light' that used the popular
degenerate configuration described in \cite{30}.

{\bf 8. Saturable absorber in the rotating frame}
 We have already mentioned that optically pumped atoms in zero magnetic field may serve as an
 adequate model of the saturable absorber. In these case the absorption saturation
of the system corresponds, depending on the light polarization, to
either orientation, alignment, or combination thereof. In the presence of an
external magnetic field, the light-induced ordering of atomic moments appears to
be hampered by their precession. However, as is known (see, e.g., \cite{17}), by
modulating the intensity or polarization of the light at the frequency of the
precession, one can realize the orientation (or alignment) in the rotating
frame. Thus one comes to numerous effects of quantum and nonlinear optics
treated in terms of coherent population trapping, double microwave-optical
resonance, beat resonances, superposition of quantum states, stimulated Raman
scattering, electromagnetically induced transparency, 'dark' polariton, etc.
(see, e.g., \cite{38}). The EIT effect is most frequently implemented in a
3-level $\Lambda$-scheme with a relatively small distance between the two low levels
formed by the Zeeman or hyperfine interaction (Fig. 1). To observe the EIT
effect, a strong pump bean is applied to one of the arms of the $\Lambda$-scheme
(with the unpopulated low level) and to the other armm, a weak probe beam. The pulse
of this probe light is used to observe the effect of the group velocity
reduction. The effect is detected when the frequency difference between the two
fields is exactly equal to that of the transition between the two low
levels $|1\rangle$  and $|2\rangle$. "Exactly" means to within the width of this transition,
which determines spectral width of the 'transparency window' and may be
extremely small. From the narrowness of this transparency window it is concluded
that the spectral dependence of the dispersion of the medium at the frequency of
the probe beam (at resonance) may be extremely steep, and that the group velocity
(due to the {\it normal} spectral dependence of the dispersion), is extremely low.

The experiments on 'slow light' frequently use phase-correlated light beams
obtained either by shifting the frequency of a single laser source or by
phase-locking their difference frequency. Under these conditions, the fact of
observation of the EIT resonance upon scanning of the {\it frequency difference}
between the beams (i.e., the fact that the resonance is observed in the
low-frequency domain of the transition $|1\rangle$ - $|2\rangle$) does not mean that such a
narrow resonance is present in the {\it optical} spectrum in the vicinity of the
resonance $|1\rangle$ - $|3\rangle$. Moreover, when the spectral width of the laser source
substantially exceeds that of the EIT resonance (which is frequently the case),
such a narrow resonance in the optical range cannot exist in principle. In other
words, under these conditions, {\it the medium is probed by the difference
frequency}, whose high monochromaticity provides a small width for the instrumental
function of this spectroscopic technique in the relevant frequency range.

This circumstance is often overlooked in theoretical studies of the EIT effect
performed in terms of monochromatic waves. In this case, scanning one of the
frequencies is evidently identical to scanning the difference frequency, and the
question about phase succession of the beams loses its meaning. At the same time,
this circumstance is, in our opinion, highly important for interpretation of the
experiments on 'slow light', because in a considerable number of these studies,
the narrow resonances of the effects of EIT and coherent population oscillations
are observed under conditions of scanning of the difference frequency. In this
case, the conclusion about a narrow resonance in the optical spectrum can be
made only when reliable information about spectral width of the light beams is
available. This remark may be addressed both to most papers on 'hole burning'
under conditions of coherent population oscillations \cite{19,20,21,22,23,24,26}
and to a number of papers on the EIT-based 'slow light' \cite{18,33,39,40}.

{\bf 9. On the intensity spectra}
 We have already pointed out that the distortion of a smooth light pulse transmitted
through a saturable absorber is reduced to a pure shift provided its temporal width exceeds
the population relaxation time. A similar requirement reformulated into the language of
frequencies is imposed upon the probe pulse of 'slow light'  for EIT effects or
coherent population oscillations, when the pulse spectrum should fit into the narrow
transparency window. The difference is that, in the first case, we deal not with the
spectrum of optical signal but rather with its {\it intensity spectrum} (whose amplitude,
in particular, at optical frequency vanishes). The {\it intensity spectrum} of the optical
pulse can be narrowed indeed by changing its duration, but the optical spectrum of the pulse
can be varied by changing its duration only if the pulse is transform-limited.
This circumstance is often overlooked, and the spectral width of an optical pulse is implied
to be entirely controlled by
the pulse duration (see, e.g., \cite{18}).

{\bf 10. Spatial aspect} When using gaseous nonlinear media, with moving
elementary carriers of the optical nonlinearity (in particular, the optically
pumped atoms), the light-induced anisotropy may come out beyond the beam
dimensions, spreading over the volume occupied by the system.. The degree of this
spreading evidently depends on the ratio between the
relaxation and diffusion parameters of the atomic medium. Specifically, in the
paraffin-coated 'vacuum' cells (with no buffer gas), a narrow light beam, a few
millimeter in diameter, is able to completely orient (or align) the atoms in a
cell several cm in size. The light-induced anisotropy (or coherence) may, in this case, be successfully
detected by a probe beam passing through any part of the cell.
In the cells with buffer gases, this fairly obvious effect was observed, in
various modifications, beginning from 1967 (see, e.g., \cite{41,42,43}). This is
why we cannot agree with the authors of \cite{44} who consider this effect as
''significantly expanding the capabilities of the quantum
information storage technique''.

\section{DISCUSSION AND CONCLUDING REMARKS}

In these notes, we have considered, in a qualitative way, some simple effects
observable in the media with nonlinear absorption, which are phenomenologically
close to some manifestations of 'slow light', but have nothing to do either with
a narrow spectral dip in the absorption spectrum of the medium, or with a steep
dispersion of its refractive index, or with the 'slow light' proper. When
considering all these phenomena in the framework of the model of saturable
absorber, we have also touched on a non-degenerate $\Lambda$-scheme by passing to the
rotating coordinate frame, when the 'diagonal' relaxation of populations is replaced by the
relaxation of coherence, and the effects of {\it incoherent} nonlinear
optics (saturation effects), to which our discussion was originally supposed to
be restricted, are transformed into the effects of {\it coherent} nonlinear
optics. However, this does not change the essence of the matter.

In all the
cases considered above, the retarded response of the medium is a result of
boundedness of its frequency passband. In the 'slow light' effects, the
extreme narrowness of transparency window gives rise to similar limitation in the
transmission bandwith. This accounts for the phenomenological similarity of
these essentially different effects. The difference is that, in one case, this
'narrow' band is localized in the range of optical frequencies, whereas, in the
other, in the range of much lower (e.g., zero) frequencies. However, when
detecting the {intensity spectrum} of the transmitted light, localized at low
frequencies, this difference vanishes. To really observe 'slow light' with ultralow
group velocity  in the sense of Eq. (1), one has to necessarily provide the highest
frequency stability and spectral narrowness of the light beams used in the experiment.

It seems evident that the nonlinear effects considered above are much less
exacting to the properties of the nonlinear medium and light beams and can be
observed much easier. These are trivial and rather universal effects of
nonlinear optics that should be taken into account by the experimentalist {\it
first  and foremost}. However, in the studies on 'slow light', these trivial
possibilities, as a rule, are not considered at all. In fact, as was already mentioned,
many papers do not contain highly important information about spectral widths of
the light beams. The experiments are frequently performed with phase-correlated
light beams with a well defined difference frequency, but with poorly defined
optical frequencies and, hence, with a fundamentally uncertain position of the
'transparency window'. The effects of slowing down and storage of light are
frequently demonstrated using the 'degenerate' $\Lambda$-scheme, which perfectly models
the polarization saturable absorber and is unsuitable for demonstrating specific
properties of the EIT effect (at least for totally coherent beams). A delay of
the light pulse in the medium (in different experimental arrangements) is always
ascribed to a change of its group velocity related to a steep dispersion of the
refractive index. The group velocity is always calculated by dividing the length
of the medium by the delay time of the pulse maximum. It is evident that the
quantity with the dimensions of velocity obtained in this way may have nothing
in common with the group velocity of light in the medium. This universal
practice of measuring the light group velocity in some cases becomes obviously
self-exposing. In particular, in \cite{26}, a sluggishness of the photoresponse
(in the range of seconds) of bacteriorhodopsin molecules in a polymer film
is ascribed to the spectral hole-burning effect under conditions of coherent
population oscillations, and the group velocity measured in the standard way was
found to be 0.091 mm/s. There is no question that the achievement of this kind can
be set up {\it ad infinitum}. An even more striking example is the experiment
\cite{45}, in which a retarded holographic
reconstruction of the writing light beam in a photorefractive crystal was detected. In this
study, the group velocity calculated in the same universal way was as low as 0.025 cm/s
(with the shape and amplitude of the pulse at the exit of the crystal modified
drastically). Of course, this effect (as well as the holography proper) may be
referred to as 'slow' or 'stopped' light, but is there any sense in these
terminological manipulations? The authors frequently ignore the fact that the
standard formula for the light group velocity (1) is applicable only to a {\it
linear, optically transparent} medium and cannot be applied to a {\it nonlinear
absorber} , when the speed of motion of the pulse peak cannot be identified with
the {\it group velocity of light}. At the same time, the studies on 'slow light'
frequently do not pay sufficient attention to the experimental facts that
contradict canonical models of the effect. Among them may be mentioned, in
particular, the dependence of the 'released' pulse shape on the pump beam
intensity \cite{34}, the admissible nonadiabaticity of switching of the pump
beam \cite{44}, and an ideal agreement of the results of the experiments on 'slow
light' under conditions of 'coherent population oscillations' with the
prediction of the trivial model of saturable absorber (see, \cite{15,25}).

In our opinion, to observe the 'slow light' effect, in the sense assigned to it
by the authors of this term \cite{7}, it is not enough to demonstrate a delay of
the light pulse maximum in the medium, or phase delay of the amplitude
modulation of the probe light, or polarization memory of a photochromic medium.
Furthermore, specificity of this effect allows one fairly easily to distinguish it from
standard manifestations of properties of a nonlinear absorber. Stress once
again that the difference between the pulse distortion in the saturable absorber
(including the polarization scheme considered above) and dispersion-related
pulse retardation is of {\it physical} nature and cannot be attributed to
terminological discrepancies (as sometimes claim our opponents). The dispersive
basis of the group velocity variation in the 'slow light' effects is known to be
revealed not only in specific features of their spectral behavior but also in a
dramatic {\it spatial compression} of the light pulse in the medium. This
popular image is frequently used in publications on 'slow light' (see, e.g.,
\cite{30,37}), whereas, real compression was demonstrated only in the classical
experiments with ultracold atoms \cite{46}, as well is in recent experiments
with photonic crystal waveguide structures \cite{47}. Unfortunately, the
difficulties of experimental observation of this compression usually do not
allow one to use it for diagnostics of the 'slow light' effects. As an implicit
evidence of the spatial compression of the pulse may serve the pulse delay
exceeding the pulse width. In this case, the spatial size of the light
pulse cannot exceed the size of the medium. It is implied, of course, that the
changes in the pulse shape are negligibly small. The pulse delays considerably
exceeding their width has been indeed observed in \cite{10,11}, in which the
conditions for observation of the EIT effect were satisfied, and the interpretation
of the results raises no questions.

As for the prospects of application of 'slow light' in atomic vapor for the
buffering and processing of optical signals, these hopes seem to be
substantially deflated by the narrow frequency band inherent in this effect
\cite{48}. Perhaps, more interesting, in this respect, are the experiments on
the light slowing down in optical fibers under conditions of stimulated Raman or
Brillouin scattering \cite{49}.

In these notes we did not pursue the goal to give a comprehensive
review of publications on 'slow light', and, of course, a great number of papers
devoted to these problems remained beyond the scope of our discussion. Still,
the above analysis allows us to definitely state that a considerable number of
claims on observation of 'slow light' are erroneous or, at best, groundless. The
situation is aggravated by the fact that even the most evident physical errors
of these papers systematically remain 'unnoticed', and the relevant publications
do not find critical evaluation in the literature, holding their high citation
level. It is noteworthy also that all the papers involved into this trend,
regardless of their scientific novelty and degree of reliability, are published
in the most prestigious journals and automatically acquire a high rating, whereas
the attempts to express, in the same journals, some doubts about correctness of
these 'achievements' meet a strong corporative resistance. This state of
affairs, in our opinion, brings damage both to the prestige of science and to
recognition of genuine achievements of this trend of physical optics.

\begin{figure}
\epsfxsize=400pt
\epsffile{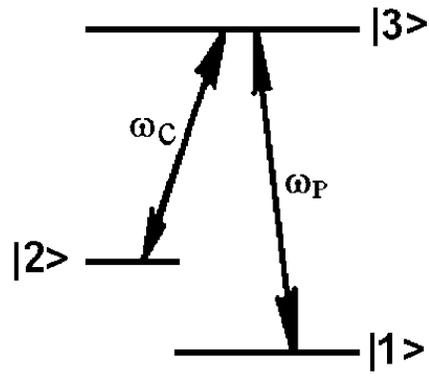}
\caption{A model for the three-level energy diagram for observation of the
electromagnetically induced transparency effect ($\Lambda$-scheme).}
\end{figure}
\newpage

\begin{figure}
\epsfxsize=400pt
\epsffile{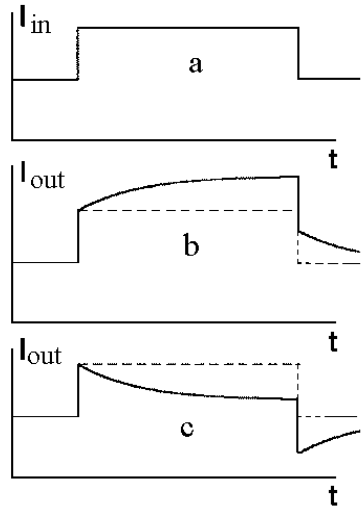}
\caption{ Schematic depiction of time dependence of the light intensity at the exit of a
usual (b) and inverse (c) saturable absorber for a step-wise change of the
intensity at the entrance (a).}
\end{figure}
\newpage

\begin{figure}
\epsfxsize=400pt
\epsffile{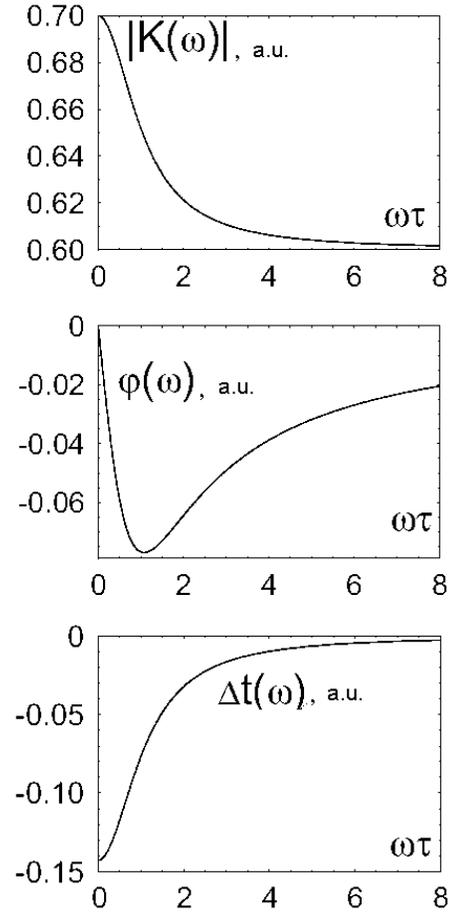}
\caption{Typical frequency variations of the amplitude $|K(\omega)|$, phase
$\varphi(\omega)$, and time delay $\Delta t (\omega)$ of intensity oscillations
of the light beam passed through a saturable (bleachable) absorber.}
\end{figure}
\newpage

\begin{figure}
\epsfxsize=400pt
\epsffile{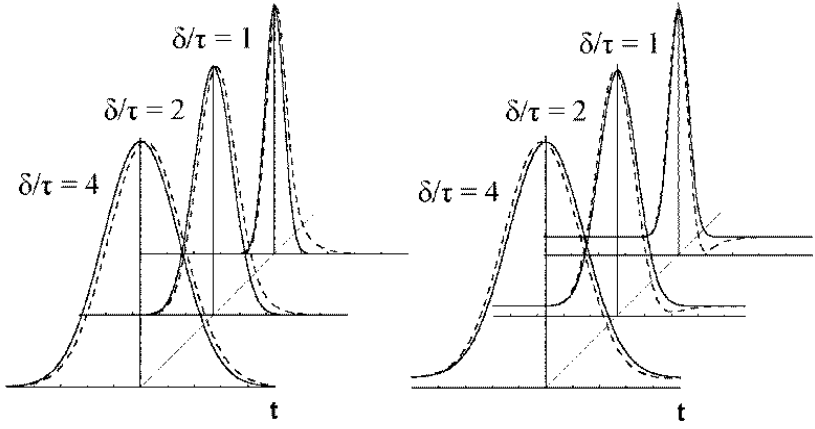}
\caption{Normalized Gaussian pulses $I(t) \sim \exp -(t/\delta)^2$ propagating
through a usual (a) and inverse (b) saturable absorber for different ratios of
the pulse width $\delta$ and the relaxation time of the absorber $\tau$:
$\delta/\tau$ = 1 (1), 2 (2) and 4 (3). Solid lines - pulses at the entrance,
dashed lines - pulses at the exit [15].}
\end{figure}
\newpage

\begin{figure}
\epsfxsize=400pt
\epsffile{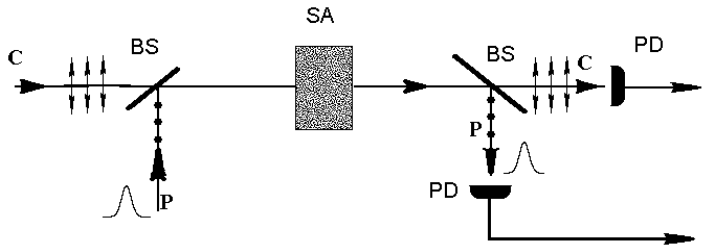}
\caption{The pump-probe configuration, popular in the 'slow light' experiments
with coherent orthogonally polarized beams from a single  source
('degenerate' $\Lambda$-scheme). Ñ - pump (coupling) beam, Ð - probe beam, SA -
saturable absorber, BS - polarization beamsplitter, PD - photodetector.}
\end{figure}
\newpage

\begin{figure}
\epsfxsize=400pt
\epsffile{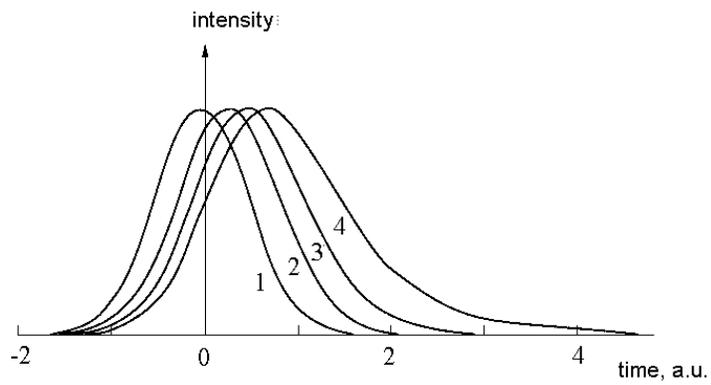}
\caption{The shape of the 'probe' pulse (in the arrangement shown in Fig. 5) at
the entrance (1) and at the exit (2 - 4) of the medium for different ratios of
the pulse width $\delta$ and relaxation time of the absorber $\tau$: $\tau =
\delta/2 (2), \delta (3)$, and $ 2\delta (4)$ [15].}
\end{figure}
\newpage

\begin{figure}
\epsfxsize=400pt
\epsffile{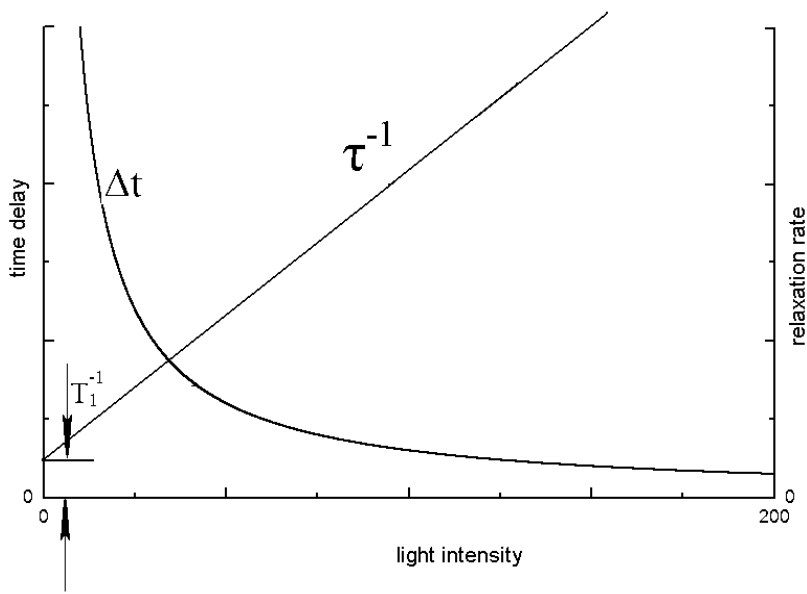}
\caption{Standard dependence of the pulse (or intensity oscillation) delay time
at the exit of a nonlinear absorber and of the absorption relaxation rate on
light intensity.}
\end{figure}
\newpage

\begin{figure}
\epsfxsize=400pt
\epsffile{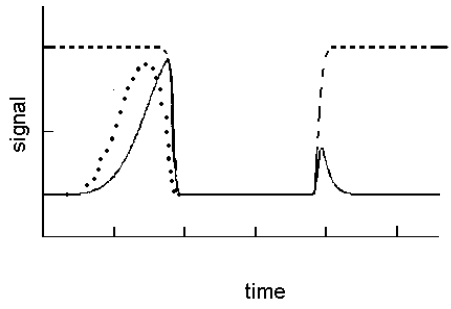}
\caption{Standard time dependence of the 'probe' beam intensity (solid line) in
demonstrations of 'stopped light' in the arrangement shown in Fig. 5. Dotted line -
intensity of the 'probe' beam at the entrance, dashed line - intensity of the
pump beam at the entrance.}
\end{figure}


\end{document}